# Web-Based Multi-View Visualizations for Aggregated Statistics

Daniel Hienert, Benjamin Zapilko, Philipp Schaer and Brigitte Mathiak
GESIS – Leibniz Institute for the Social Sciences
Lennéstr. 30, 53113 Bonn, Germany
+49 - (0)228 - 2281 - 0

{daniel.hienert, benjamin.zapilko, philipp.schaer, brigitte.mathiak}@gesis.org

## ABSTRACT
With the rise of the open data movement a lot of statistical data has been made publicly available by governments, statistical offices and other organizations. First efforts to visualize are made by the data providers themselves. Data aggregators go a step beyond: they collect data from different open data repositories and make them comparable by providing data sets from different providers and showing different statistics in the same chart. Another approach is to visualize two different indicators in a scatter plot or on a map. The integration of several data sets in one graph can have several drawbacks: different scales and units are mixed, the graph gets visually cluttered and one cannot easily distinguish between different indicators. Our approach marks a combination of (1) the integration of live data from different data sources, (2) presenting different indicators in coordinated visualizations and (3) allows adding user visualizations to enrich official statistics with personal data. Each indicator gets its own visualization, which fits best for the individual indicator in case of visualization type, scale, unit etc. The different visualizations are linked, so that related items can easily be identified by using mouse over effects on data items.

## Categories and Subject Descriptors
H.3.5 [**Information Storage and Retrieval**]: Online Information Services

## 1. INTRODUCTION
More and more governments, statistical offices and other organizations offer their statistical data as Open Data or Linked Open Data. The range of data covers many areas from the world's oil consumption to fishery production in Mediterranean countries. Statistical data and in particular time series can be very complex in aspects like dimensions, structure, size and format. To interpret and analyze these indicators easier they have to be visualized in order to identify trends, minima, maxima etc.

First efforts to visualize this data are made by the data providers themselves. For example, Eurostat provides online graphics like bar charts or maps, which show individual indicators in simple understandable visualizations on a time and spatial distribution. Data aggregators like datamarket.com go a step beyond. They collect data from different open data repositories and make them comparable by providing the possibility to choose data sets from different providers and showing the different statistics in the same line chart. This way, different indicators with similar temporal and spatial information can be compared in one integrated visualization. Another solution, Gapminder, depicts two different indicators in a scatter plot.

The existing solutions have several drawbacks: using a single visualization provides an integrated view, but makes it harder to distinguish between indicators. They are often only color-coded, so that the identification has to be done with the help of the legend or by a mouse-over. If too many indicators are chosen, the graphic gets visually cluttered and one cannot distinguish between the time series. Different visualization types provide more possibilities to choose an adequate visualization for the individual indicator, which makes it easier to use visualization type characteristics to detect individual data items. Regarding different units and ranges of data the different indicators may have different scales (i.e. percentages, absolute numbers), which complicates comparisons of the indicators.

Our approach marks a combination of (1) the integration of live data, (2) presenting different data indicators in coordinated visualizations and (3) allows adding user visualizations to enrich official statistics with personal data. Each indicator gets its own visualization, which fits best for the individual indicator in case of visualization type and scale. For each indicator the visualization type and individual data of countries can be chosen. The different visualizations are connected by a link, so that hovering with the mouse over a data item highlights related items in all other visualizations and shows their values.

In the next sections, first, we will present related work and ideas. In section three, we will discuss state-of-the-art in visualizing official statistics. In section four, we will give a brief overview on the capabilities of our tool and will present the approach of coordinated statistics in our system. We will proceed in section five with a case study and will discuss advantages and disadvantages of different approaches to compare statistics. We conclude with some final remarks and future work we have planned.

## 2. RELATED WORK
In this section we present work and ideas which are related to visualization on the web, linked visualization and the coordinated view approach.

IBM Many Eyes [17], Data360 [2], StatCrunch [19] and Socrata [16] are online tools for sharing data and visualizations on the web. The user can upload data, choose a visualization type and create a visualization that can be viewed and commented by the community. By integrating an HTML snippet the visualization can be embedded on other sites or blogs. The underlying data set can be reused by other users to build their own visualizations. Heer et al. [6] give an overview of online visualization tools, their functionality and impacts.

Wang Baldonado et al. [18] set up a model for coordinated multiple views and provide guidelines for not disrupting the positive effect through increased complexity. The main idea is that data in different views can be linked. If data is selected in one view, it is also highlighted in other views (brushing-and-linking). North & Shneiderman provide an alternative visualization model which is based on the relational data model [12]. The system Snap [13] is an implementation of this model. It allows the user to select databases and assign visualizations. In a second step, the user can then connect different visualizations and generate coordinated visualizations. Highlighting or other actions are coordinated between the different views. VisGets [3] uses different visualizations to show and filter retrieved web resources in several dimensions like time, location and topic. Based on the concept of dynamic queries, results can interactively be filtered by manipulating the visualizations. VisGets also implements coordinated interactivity. Hovering with the mouse over a visual element highlights all related elements in the visualizations and in the result list. The new introduced approach of Weighted Brushing is used to highlight strongly related items more than weakly related ones. Tableau [10] or Spotfire [1] are solutions which allow the creation of dashboards with different visualizations. Dashiki [11] is a wiki-based collaborative platform for creating visualization dashboards. Users can integrate visualizations that contain live connections to data sources. Data sets are embedded into data pages by a special markup, via Copy & Paste from spreadsheets or by a URL. Live data is dynamically fetched and stripped from formatting tags, so the user can wrap the content with needed markup. Dashiki uses a simple technical approach for coordinated selecting among multiple views. Simple attribute-value pairs are propagated to all visualizations via JSON format. Exhibit [8] is a lightweight framework for easy publishing of structured data on the web. Users can import data via JSON, which is presented on the web page in different views including maps, tables, thumbnails and timelines. Scientists from different domains often work with statistical software like STATA, SPSS or the R Project. These packages contain basic graphical functionality to create different charts, but data must be in their own format and cannot be loaded live into the application. For the comparison of multiple statistics only scatter plots can be used.

## 3. VISUALIZING OFFICIAL STATISTICS

### 3.1 By Data Providers

Data providers offer their own tools for visualizing their statistical data. Common visualization types are bar/line/area/pie charts and scatter plots or world/country maps. We show the visualization of statistics by data providers by the examples of Eurostat [4] and World Bank [20], whose data are integrated in our tool.

In the Eurostat Table, Graphs and Maps Interface (TGM) users can choose between a table, several graphs and a map view. The table shows the individual values resolved by country and time. In the graph view users can choose between several visualization types like bar, line, pie charts and scatter plots and can define temporal and spatial filters on the data set. For example a bar chart can be made for the economic activity of all countries of the European Union in 2009. The third view shows a map of Europe, where different colored countries symbolize different value ranges for one time unit. This view can also be customized.

The same threefold splitting is used by the World Bank, offering table, map and a line chart view. To allow a smarter browsing and visualizing they offer a timeline on the top and bottom of the data table that allows choosing from different time classes. This allows in the map to view aggregated data for example from 2005-2009. All views are fairly customizable but are optimized for a quick overview.

Most tools by data providers offer basic possibilities to visualize their data. The main disadvantage is that only their own data can be displayed and no data from other sources can be integrated for comparison.

### 3.2 Comparing Statistics

DataMarket.com integrates several important open data providers like the UN, World Bank, Eurostat, Gapminder Foundation etc. with about 100 million time series. Users can choose from a large catalogue of available time series classified by data providers. Because most data sets contain a large number of dimensions, users have to choose dimensions to be visualized from categories like country/area, time and from different indicators. Data is displayed in a line chart. More data sets can be included, therefore users can choose a new data set from a pop-up window and add it to the list. Again several dimensions from the data set have to be chosen and can then be visualized as a second line in the chart.

An alternative to the standard visualizations for statistical data is Gapminder [15]. Google acquired Gapminder in 2007 and integrated the Trendalyzer software in their Charts API. The Gapminder software has a chart and a map view. The graph view depicts three different indicators in a scatter plot, two on the x- and y-axis and a third is symbolized by the size of bubbles. Temporal information is used to provide an animation, spatial information is shown in a small map. Different indicators for the axis can be chosen by clicking on the axis' label and choosing an indicator in the hierarchic menu. Gapminder right now offers about 500 different indicators. Because both axes are used for the values of their respective indicator, time information can be used to provide an animation. Pressing the Play-Button animates the graph from start to end time, so that the development of one or several countries or areas can be followed depending on the indicators. Spatial information is represented three times: once in the color of the bubbles, on a small world map that shows the region and in the legend. Chart, map and legend are coordinated to highlight the chosen country. In the map view the same functionality is provided, but the view allows only one indicator to be represented in the bubble size.

### 3.3 Characteristics

The integration of heterogeneous open data sources and the integration of different statistics for visualization have certain characteristics: (1) different data sources and formats, (2) actuality, (3) dimensionality, (4) size, (5) the individual view and (6) the integrated view.

Data can derive from a variety of sources and in a variety of formats, which must be converted to a tool's specific one. For example Eurostat data can be accessed via tab separated values (TSV) and semantic web technology like RDF [9] and SPARQL [14]. World Bank data is offered in Excel and comma separated values (CSV). The internal format can also differ very much: all indicators can be listed in one table or are distributed over several tables. Rows and columns can be reversed and individual values can be formatted differently, e.g. for zero values and no values. However, with static data files a complete data update can be very time consuming. A SPARQL query is more complex, but guarantees that only data is retrieved, which is needed for the current view. The data is always up to date, since it is mostly queried from the original data source. If the output format can be adapted to TXT or CSV the processing is simple, otherwise complex RDF documents must be parsed and file and transmission size can be much higher.

An individual statistic can contain several dimensions like different indicators, seasonal adjustments, units, age classes etc. This can make data processing very complex and results in large file sizes. Even if only one item is selected for each dimension class, data size can be very large. Statistical data sets integrated in our tool contain up to 240 areas/countries and a temporal coverage from 1800 till today, partly on a monthly or quarterly basis. Because one data set can contain multiple dimensions, individual values for each dimension class has to be selected in advance, otherwise visualizations visually clutter very much. To select a reasonable selection of countries and time periods for the visualization, temporal and spatial information must be easily selectable in the user interface.

With the integrated view of several indicators in one view several difficulties are added: (1) indicators often have different units, (2) indicators often have a different scaling, (3) one integrated line chart can visually clutter very quickly and different indicators can hardly be distinguished and (4) the visualization type may not always be appropriate for every statistic or representation could be improved by using other visualization types.

## 4. STATISTICS IN VIZGR

Vizgr [7] is a visualization platform on the web where users can create own visualizations, connect them to other resources, view statistical visualizations and can combine different visualizations on one page. The tool offers basic visualization methods, like graphs, tag clouds, maps and time lines. But unlike normal data visualizations, these can be re-used, connected to each other and to websites. Vizgr offers a simple opportunity to combine diverse data structures, such as geo-locations and networks, with each other by a mouse click. In this section we will show the integration of open data sources in our system, the display and combination of these data with other statistics and user visualizations.

### 4.1 Data Sets

In addition to the uploaded data from users the system integrates open data sets from Eurostat, The World Bank, Gapminder and the European System of Social Indicators.

#### 4.1.1 Eurostat

Eurostat is the statistical office of the European Union. They offer over 5000 time series with detailed statistics from all European countries. Data is provided by the national statistical offices and is harmonized by Eurostat in terms of statistical definitions and methods. All data is publically accessible in different formats like TSV, DFT and SDMX and updated twice a day. There are a variety of topics, ranging from convergence criteria of the EU to health issues. We use the Eurostat wrapper [5] to integrate the data in our tool. The wrapper offers all data sets in RDF notation and data can be queried live with a SPARQL endpoint. Eurostat data files for individual indicators can have a size of several Megabytes. This can complicate and slow down the live processing over the internet and the immediate visualization. The use of a SPARQL endpoint makes it possible to query only a part of dimensions, countries and time periods to keep the transfer size small. This way, data can be queried live and is always up to date.

#### 4.1.2 The World Bank

The World Bank provides access to over 2000 time series. The data catalogue includes data sets to World Development Indicators, Global Development Finance, Africa Development Indicators, Education Statistics, Gender Statistics, Health Nutrition and Population Statistics and Millennium Development Goals. Data is offered in standard formats like CSV and Excel and can also be accessed via an API. Because data updates are only published at longer time intervals we integrated all data in our system to be able to offer fast visualizations.

#### 4.1.3 Gapminder

The Gapminder Foundation provides about 500 indicators from different open data providers. Some indicators are compiled by Gapminder and include various sources. If available, data covers all countries and territories of the world. Data sets come from categories like economy, society, education, energy, environment, health, infrastructure, population and work. Raw data is stored online in Google Spreadsheets and can be queried live in different formats like Excel and CSV format.

#### 4.1.4 European System of Social Indicators

GESIS – Leibniz Institute for the Social Sciences offers an indicator system for a continuous monitoring of individual and societal well-being across Europe. Information on the social situation and development in society, i.e. demographic developments, the development of wealth and quality of life, the distribution of poverty and wealth, or the realization of gender equality and education are important for policy-making and social discussions. The European System of Social Indicators (EUSI) contains over 600 indicators from 1980 to present and covers 27 EU member countries, Norway, Switzerland, as well as Japan and the U.S as major reference countries. The time series cover the following life domains: Population, Household and Family, Labor Market and Working Conditions, Housing, Education and Vocational Training, Income, Standard of Living and Consumption Patterns, Health, Environment, Crime and Public Safety and Total Life Situation. We integrated Excel dumps of this data in our system.

### 4.2 Visualizing Data Sets

Data sets from Eurostat, World Bank, Gapminder or EUSI can be selected from the gallery and will then be shown on an individual page. Depending on the data provider and the complexity of the data set, different dimensions of the individual data set can be chosen from select boxes. Common dimensions are for example different indicators, seasonal adjustments, units, age classes etc. Eurostat data sets for example can include five or more dimensions, each with multiple items. Simple select boxes allow the easy selection. By using the first entry of every dimension for visualization the user is not forced to select several dimensions in advance and then be able to view the visualization. Instead the tool provides an initial visualization and the user has the possibility to play around with different settings.

A line chart is used as visualization type, but users can choose other types like bar, area, pie chart or scatter plot. Time information is on the x-axis, values on the y-axis. Countries appear in the legend and are similar color-coded as lines, bars etc. Up to 40 countries are initially shown in the chart. The user can select different countries directly from the legend. For example by clicking on Germany and France, one can compare actual monthly consumer data with two clicks. Data sets can contain over one hundred different countries. The legend is scrollable, so that a large number of countries can be accessed and selected easily within the visualization. Most tools like the ones from Eurostat or Datamarket need to choose time and country information in advance and to create the visualization again.

## 4.3 Adding Other Statistics or User Visualizations

A critical point in the user workflow is the process of choosing a data set from a large pool. All together our prototype offers already over 7000 time series and a set of user visualizations the user can choose from. Tools like Gapminder offer a hierarchical access, but by providing data sets from heterogeneous sources the categorization is hardly possible. We created a solution that combines the selection by data provider, a search box and browsing facilities (figure 1).

Further visualizations in our tool can easily be added to the page by clicking *Add Visualization*. Users can choose data sets from the categories User visualizations, Eurostat, World Bank, Gapminder or EUSI. Available data sets are shown, can be browsed and easily be added to the page by clicking the button *Add*. With the search box users can search for keywords in titles. By adding a visualization, a scaled version of the original visualization is shown beside the added one. This way, both visualizations fit on the screen and can be compared more easily. Added visualizations have the same interactivity to choose dimensions, visualization type and to select countries.

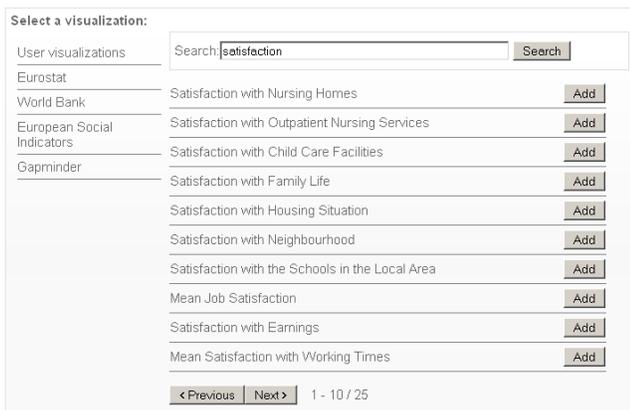

**Figure 1. Adding further visualizations to the dashboard.**

## 4.4 Coordination between Statistics and User Visualizations

Coordinated views in statistics are possible because data sets contain similar temporal and spatial information. We choose combined time and space information to create the linking. This way, values in different indicators can be compared for the same time and country. For example, hovering with the mouse over a value in statistic A for Germany in 2007 highlights all values for the combination Germany/2007 in all other visualizations. This can be achieved by using Vizgr's capabilities to link different data sets. Links between visualizations are defined by creating a rule set for every combination of country and time.

Coordination between user-generated visualizations and statistics is achieved by two approaches. If user-generated visualizations contain data that is structured similarly to statistics or contain keywords, which are used in statistics, the mapping can be done automatically. For example, if a map with European countries and places are labeled corresponding to labels in the statistical data, then hovering with the mouse over a data item of Germany also highlights the place on the map. If the visualization from the user contains no similar keywords in data items, then coordination can be achieved by mapping data of two visualizations with the Mapping Editor that is provided in Vizgr. For details on the linking approach and the Mapping Editor see our previous work on Vizgr [7].

## 5. ANALYZING STATISTICS WITH COORDINATED VIEWS

### 5.1 Official Statistics

In the following case study we will present how statistics from different data sources can be added to one page and are automatically connected. We will further add user visualizations and will see how coordinated visualizations can be used to analyze several indicators.

As a first data set we choose the *GDP per Capita (current US$)* from the World Bank pool. The Gross Domestic Product (GDP) indicates the market value of all goods and services produced within a country in one year divided by midyear population. It is often used as an indicator of a country's standard of living. The data set contains the GDP for 237 areas and countries in US$. The visualization is displayed in large size at the top of the page. Users can select different visualization types like bar/line/area/pie chart and scatter plot. Because the data set contains many attributes users have to choose countries of interest. By selecting different countries from the legend the individual values and trends over time can be compared for the same indicator in the same chart, analogous to existing tools. For example we can compare the GDP for the United States and the United Kingdom by selecting both countries from the legend. We see a steadily increasing curve for the United States from 1960 to 2008 and a small loss from 2008 to 2009 resulting from the financial crisis. The curve of the United Kingdom contains several local minima.

To complement the GDP with a second indicator, we click on *Add Visualization* and search for the keywords *life expectancy*. Eurostat and The World Bank offer data sets, we choose the data set *Life expectancy at birth, total (years)*. This data set indicates the number of years a newborn would live, if prevailing patterns of mortality at the time of his birth would stay the same throughout his life. So, it describes life expectancy for the time of birth in years for both genders. The new visualization appears below the original visualization and beside a smaller version of the original visualization to make comparison easier. We can now compare GDP and Life Expectancy in Europe and Africa by selecting these areas in both visualizations. We can see a great difference in the GDP graphic between Africa and Europe. Africa starts with a value of 151$ in 1960 and ends with 1593$ in 2008, but with a relative stable level between 1980 and 2002 with values between 650$ and 800$. In contrast the European Union starts with 904$ in 1960 and ends with 32,838$ in 2009. Life expectancy is rising for both areas, but on different levels: for Africa from 42 years to 54 years, for Europe from 69 years to 79 years. Both visualizations are coordinated, which means, that hovering with the mouse over a data item in one visualization highlights data points and shows values for the same time/country combination in all other visualizations. For example, hovering with the mouse over the data point 1993 in Europe in the GDP chart returns a value of 15,749$, simultaneously the data point in the life expectancy graph is highlighted and shows a value of 75 years. In 2008 the GDP has doubled with 36,834$ and life expectancy has increased to 79 years. In contrast, Africa has in 1992 a GDP of 720$ and a life expectancy of 53 years and in 2008 a GDP of 1350$ and a life expectancy of 55 years.

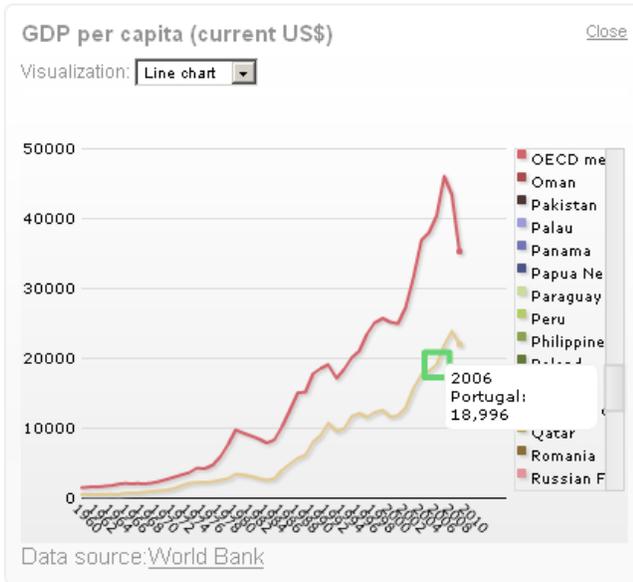
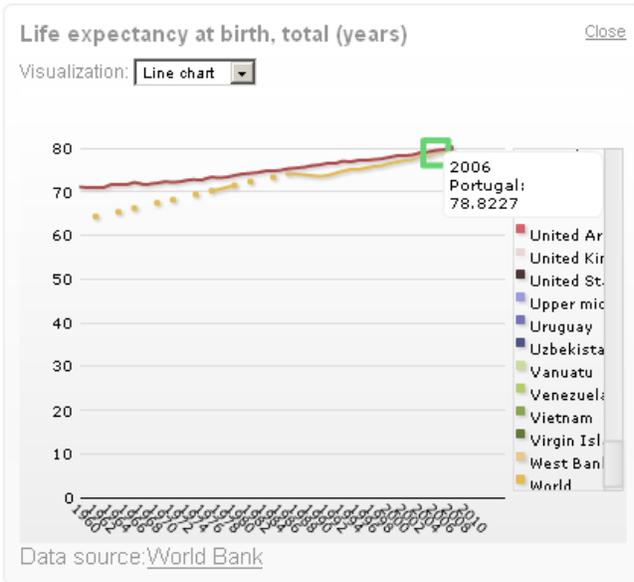
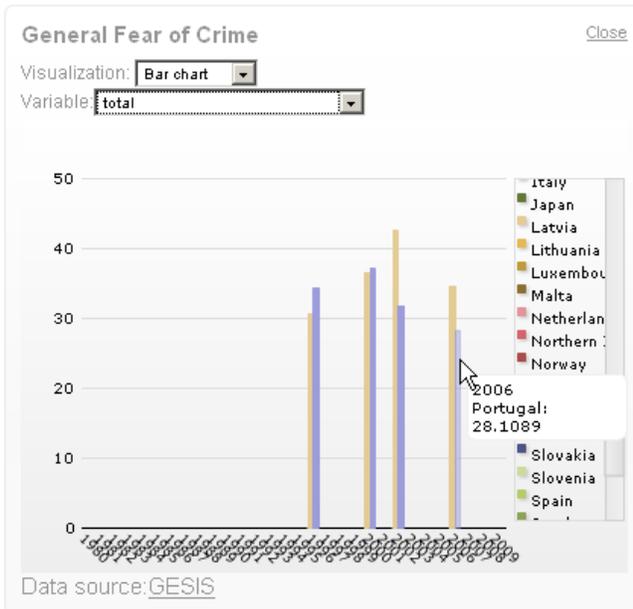
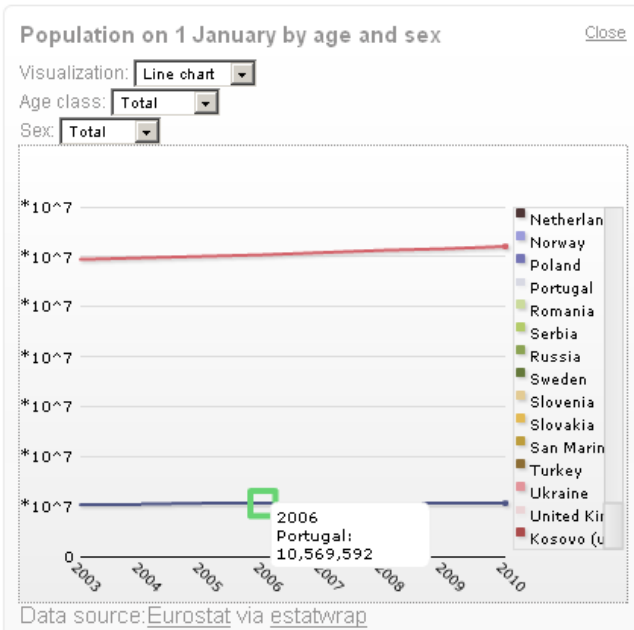

**Figure 2. Analyzing four different indicators: (1)** *GDP per capita***, (2)** *Life expectancy***, (3)** *General fear of crime* **and (4)** *Population on 1 January* **in comparison of the United Kingdom and Portugal. Hovering with the mouse over a value in the** *General Fear of Crime* **bar chart highlights related items with the same time and country information in all over charts and shows their values.**

We add a third indicator *General Fear of Crime* from the European Social Indicators as an indicator for subjective perception and evaluation of public safety. It indicates the "Percentage of people who feel very or a bit unsafe walking alone in the area where they live" from the Eurobarometer survey for Europe and from the Gallup Organization for the US. This indicator in contrast to the preceding ones is a subjective well-being indicator, whose data base is a survey of the public. We can find for example that in the US the perceived fear of crime has increased from 30% to 35% from 2001 to 2002, also the GDP has increased from 35,898$ to 36,796$ and life expectancy has increased from 77.0341 years to 77.2366 years. In contrast in Germany the fear of crime decreased from 1996 to 2000 from 39.4% to 35.1%, while the GDP decreased strongly from 29,769$ to 23,114$ and life expectancy increases slightly from 76.6732 years to 77.9268 years.

For completeness we can add a fourth indicator that shows population size. We choose the data set *Population size on 1. January by age and sex* from the Eurostat pool. We can now for example compare GDP, life expectancy, fear of crime and population size for several European countries (figure 2).

## 5.2 User Visualizations

Another possibility is to add user-generated visualizations. For example, charts with aggregated data or other visualization types like maps or a time lines. For example, we can add a map with European countries. Users can create this kind of map easily in our tool by entering locations with title, details and address. Automatic coordination can be achieved most simply, if places are labeled analogous to labels in statistical data. This way, hovering with the mouse over a country in the map highlights data items in the chart or vice versa (figure 3). In this way, a user-generated map can be used as a filter for all statistical charts or the other round hovering over a data item in a chart shows the appropriate country on the map. This is a similar approach as in other solutions, where map/charts combinations are used. But in our case we can customize the map to special needs and one map can be used to highlight values in several statistical charts. For example, some Eurostat indicators have a focus on Mediterranean countries. We can easily create a map, which has a better focus on these countries and also options for zooming and panning in contrast to a fixed non-interactive map of Europe.

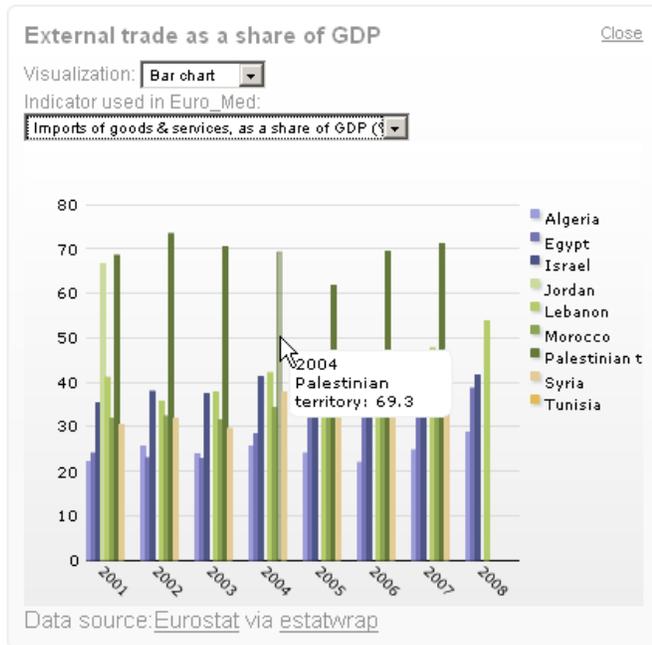
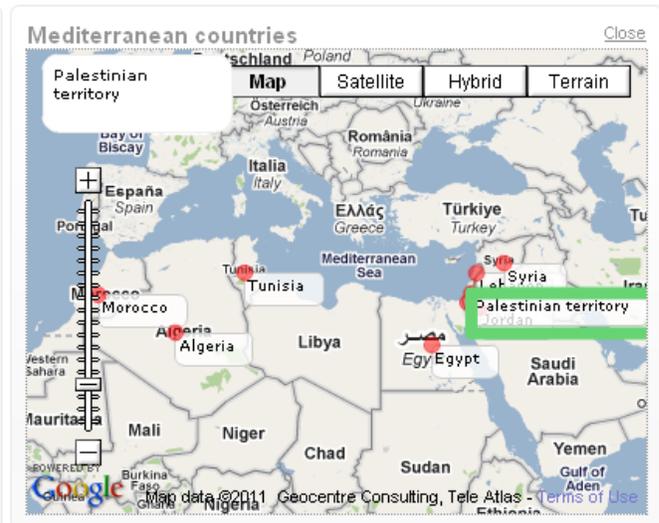

**Figure 3.** Adding a user-created map of Mediterranean countries to a Eurostat statistic of *External trade as a share of GDP* for these countries. Hovering with the mouse over a data point in the chart highlights the country on the map. The other way round, hovering over the country in the map would highlight data points in the chart and show their values.

Alternatively, we could add a user-generated time line of historical events that are in the context of recessions in the United Kingdom. User can create a time line by entering start-, end date and details of historical events like the early 1980s recession, the early 1990s recession and the late 2000s recession. Because here no direct similarities between historical events and chart values exist we can map data items on a visual level with the Mapping Editor, which is included in our tool. Events can be mapped to maxima and minima of the GDP chart just by clicking the event in the time line and a region in the chart. As a result, we have created a mapping as a basis for coordinated views. Hovering with the mouse over a region in the GDP chart highlights the corresponding event and the user can read details. The other way round, hovering over events highlights corresponding regions in the chart.

## 6. DISCUSSION

We will now discuss in detail advantages and disadvantages of the (1) integrated view of statistics in graphs (2) the comparison in scatter plots, (3) the view on maps and (4) the coordinated view approach.

Tools from Eurostat and DataMarket offer the possibility to choose time series from different datasets and to show them in one integrated chart, i.e. in a line chart. This has the great advantage, that both time series can be compared directly in one visualization with the same units and scaling for x- and y-axis. For example, for two time series, which have the same unit and a more or less similar scaling, one can directly see differences for a certain time and country combination by seeing the distance between two data points of both lines in contrast to other distances in the graph or comparing both values in detail by checking with a mouse-over. For this case the integrated view is an ideal solution as it provides the opportunity to compare two or more indicators quickly in one chart.

But the approach has several drawbacks, which become important in real life scenarios: (1) statistics often have different units, (2) statistics often have a different scaling, (3) one integrated chart can visually clutter very quickly and different indicators can hardly be distinguished and (4) the visualization type may not always be appropriate for every indicator or representation could be improved by using other visualization types.

Statistics hardly have the same unit for indicators to compare. The range goes from absolute numbers, percentages, numbers in thousands, millions, index values (2005=100) etc. This is accompanied by different scaling from numbers between zero and one for percent values, to numbers around 100 for index values to thousands and millions for absolute numbers. This results in difficulties for comparing two different indicators with different units or scaling. For example, for two indicators, one in percent and the other in absolute numbers of thousands, comparison in one line chart will be difficult. The scaling for the y-axis will be automatically adjusted to thousands. This means the time series in percent will be very close to the x-axis and can visually hardly be compared to the second one. One solution is to apply a second unit to the y-axis. With a certain conversion factor, for example 1:1000 a second unit is assigned. This way both indicators are in a similar range and can be compared more easily. But this solution is limited to two units or scaling in one chart and the display is getting more complex and may not be intuitively understandable.

Charts can visually clutter very quickly by too many indicators and users cannot distinguish easily between different indicators. For example, election results for five parties over the last 20 years results in five lines in a line chart. For comparison, we could add indicators for personal and social satisfaction with living conditions, to find out, if these indicators have influences on election results for certain parties. The line chart now contains seven lines, which are color-coded similar to the legend. To distinguish between indicators, the user has either to compare the color in the legend and of the line or use a mouse-over to see which line represents which indicator. Mixed with different scaling and units it gets hard to compare different indicators or to analyze influences of one indicator to another.

Tools often use line charts or other charts to compare indicators. The visualization type, which is used for one indicator, may not be appropriate for another. For example, high resolution data can be displayed well in a line chart, low resolution data better in a bar chart, share data better in a pie chart and single-column or accumulated data better on a map. By the integration and comparison in one chart only one chart type can be used, even if another type could better support the presentation with its specific characteristics.

Gapminder uses scatter plots to compare two indicators. One indicator is plotted on each axis; a third can be represented by bubble size. This type of display makes it very easy to relate indicators with each other. Relations between two indicators can intuitively be read on the x- and y-axis. High values for both indicators can be found in the upper right corner, low values for both indicators in the lower left corner. The problem of different units and scaling here is mitigated somewhat because they are represented on different axes. In addition, in Gapminder, for both axes a linear or logarithmic scaling can be chosen. In the integrated view with graphs the running of values over time for different spatial units is in the foreground. In the scatter plot, the actual values for two indicators for a specific time over all spatial units are indicated. An animation can be used to add time information, so that the running of values over time can be tracked easily by the user. Spatial information in the graph itself is color-coded or can be tracked by a mouse-over, by a small map or the legend. The scatter plot provides a very good solution to compare two indicators for a specific time slot and with the help of several interactivity tools for a specific spatial unit. The comparison of two indicators over a specific time period is possibly only with the animation, so that the review of the entire running over time is lost.

Maps can be used to show one indicator for one time slot. Different shadings symbolize different value ranges for different countries. This representation has several drawbacks: data is shown only for one time slot (Gapminder again uses animation to bypass that) and representation in different shadings is not intuitive. Maps can hardly be used to compare two indicators unless by using multiple maps side by side.

The coordinated view of different indicators in several charts can avoid some mentioned drawbacks. Because for one chart only one indicator is used, the chart provides always only one unit and the optimum scaling. No mixture of scaling and units is needed, which keeps the chart simple and intuitively understandable. Visual cluttering is avoided and can further be optimized by choosing one or more spatial units for each chart. Time information keeps intact and allows the comparison of values over time. For each chart a visualization type can be chosen, for example for visualization A a line chart and for visualization B a map view. The coordination of several views helps identifying values for the same time/space combination in all different views for different indicators. By providing not only a highlighting but showing the actual value in a pop-up it is possible to easily identify all different values.

The biggest advantage is simultaneously the biggest disadvantage: the incompletely integrated presentation, which would make it possible to see differences between time series at a glance. With the coordinated view the running over time or certain values have to be compared in different charts, one must visually seek back and forth between the different charts. Coordinated visualizations require a lot of screen place. For easy comparison, visualization must be ordered next to and above each other. It can be necessary to shrink the size of visualization, which can result in worse display and poorer ergonomics.

By adding user visualizations it is possible to further customize the visualization dashboards. Users can add other charts, maps, time lines etc. to enhance official statistics with their personal processed data. This way, they can create dashboards, where visualizations can also be used to filter data or create an overview over different data types, like tabular data, events, locations etc.

# 7. CONCLUSION & FURTHER RESEARCH

We have presented an approach to analyze official statistics, well-being data and user graphics in coordinated visualizations. The data derives from heterogeneous sources and in different formats, partly live data from RDF stores, partly Excel data from local stores. All have been integrated in one tool, which makes it possible to compare different indicators in coordinated visualizations. Users can select certain data in one visualization and appropriate items based on time and country information will be highlighted in all other visualizations with the appropriate value. To enrich official statistics with personal prepared data, users can add their own created visualizations. This cannot be only tabular data and suitable graphics, but also time lines, maps and network graphs. Coordination between official statistics and user visualizations is achieved by either an automatic mapping, if data items have similar labels or by a manual mapping, where users can create own mappings between data items. Result is a dashboard with a mixture of official statistics, well-being data and user visualizations, in which correlations between different indicators can be examined more easily.

A research question in the social sciences is how far historical events have influences on certain indicators. For example, have historical events like wars, financial crisis or natural disasters influence on indicators like life satisfaction, fear or health. We plan to integrate a time line that contains historical events for a certain time period. Data derives from DBpedia, Wikipedia and other sources like the New York Times. If the user selects one data item in a statistical chart, the time line scrolls to the appropriate event for this country/time combination on the timeline. This can help identifying historical events which point to local minima and maxima in the chart.